# Requirements Engineering for Machine Learning: A Review and Reflection

Zhongyi Pei, Lin Liu, Chen Wang, Jianmin Wang
National Engineering Research Center for Big Data Software
School of Software, Tsinghua University
Beijing, China
{peizhyi, linliu, wang_chen, jimwang}@tsinghua.edu.cn

*Abstract*—Today, many industrial processes are undergoing digital transformation, which often requires the integration of well-understood domain models and state-of-the-art machine learning technology in business processes. However, requirements elicitation and design decision making about when, where and how to embed various domain models and end-to-end machine learning techniques properly into a given business workflow requires further exploration. This paper aims to provide an overview of the requirements engineering process for machine learning applications in terms of cross domain collaborations. We first review the literature on requirements engineering for machine learning, and then go through the collaborative requirements analysis process step-by-step. An example case of industrial data-driven intelligence applications is also discussed in relation to the aforementioned steps.

## I. INTRODUCTION

TODAY, the world is witnessing many successful applications of machine learning techniques, including image recognition, speech recognition, traffic prediction, self-driving cars, virtual personal assistants, buyers' preference prediction and product recommendations [1]. In recent years, there are many research efforts on understanding how the software engineering processes should response to the needs of machine learning applications, and what changes have data-intensive intelligent systems brought to requirements engineering [2].

In requirements engineering, there are growing interests in understanding various needs and aspects of machine learning application systems. Research topics of interest include the non-functional requirements elicitation and quality assurance of machine learning models and applications, especially the ones different from traditional information systems developments. For instance, performance metrics, such as precision and recall, F-measure, ROC curve, are critical acceptance criteria for the viability of specific machine learning algorithms in specific contexts, which also direct the continuous optimization of ML models. In addition, Berry discussed requirements specifications for AI applications in terms of performance measures acceptable in a given context, as a value or criteria [3]. Other well-discussed topics include the explainability of machine learning models [4], the fairness and unbiasness of predictive analysis results [5], the legal and ethical compliance requirements of ML intensive systems, etc.

There are three sub-disciplines involved, namely software requirements engineering, data and knoweldge engineering, and artificial intelligence/machine learning involved. In requirements engineering, various conceptual modeling approaches are used to elicit software system requirements and specify the expected system structure and behaviour. For instance, goal-oriented requirements modeling first represents the high-level objectives of system users and designers, and then elaborates on the success and acceptance criteria of required system by goal decomposition and refinement [6]. After fully understanding the high-level objectives, system archtecture and behavior are designed and represented as formal/semi-formal modeling specifications. For example, automata and state machine diagrams in UML and SysML diagrams [7] are provn useful in analysing reactive systems requirements, specifying domain object properties and business logics through human understandable patterns, and widely used in the domain of industrial automation and control. Besides, quality assurance to specified system behaviors and causal relationship can be conducted by formalized verifications and validations [8].

On the other hand, in many science and engineering domains, there are dominating physical or process models, such as mechanical models in mechanical engineering, chemical reaction models in chemical engineering, structural mechanics models in building and construction etc. The mathematical models are in the form of equations, directed causal networks, 3D simulations of structures or dyanmic behaviors [9], which defines the nature of the learning problem, the structure, the loss functions and hyperparameters of neural networks models and algorithms, referred to as machine learning models.

The collaboration of people with different expertise is considered a major challenge, as we need to bridging semantical gaps between different knowledge areas, integrating interdisciplinary methods and tools into a coherent process, and generating evolvable learning systems.

This paper aims to provide an overview of the collaboration among the different roles in requirements engineering for machine learning systems. We first review the literature on requirements engineering for machine learning, and then dig into what each role concerns during the collaborative

Financial Support from National key Research and Development Program Project 2021YFB1715200, and NSFC Innovation Group Project 62021002 is gratefully acknowledged.

requirement understanding and system development process. We further summarize the typical patterns for collaborations, and propose high-level guidelines for evaluation and selection of viable patterns.

The rest of the paper are structured as follows: Section II explains our research method, by which we select literature papers; Section III gives our analysis result, a brief review of related work and a summary of the general concerns and challenges of collaboration; In Section IV we propose a collaborative requirements analysis process and present one example case and the lessons learnt from actual requirements analysis; Section V concludes the paper.

## II. RESEARCH METHOD

Research on RE4ML (requirements engineering for machine learning) has attracted growing interest in recent years. In this section, we first raise the research questions, and then introduce our review method. The review protocol includes: (i) how to select the document sources; (ii) what to use as the search string; and (iii) the inclusion or exclusion criteria in this review. Following this protocol, the researchers performed a parallel search in order to identify studies that address the research questions.

### A. Research Quesions

The main research questions we aim to answer in this paper are as follows:

RQ1: What are the roles involved in engineering data-driven intelligence applications?
RQ2: What are the major areas for engineers playing different role to collaborate during requirements stage?
RQ3: What kind of support a collaborative requirements engineering for machine learning is needed?
RQ4: What are the important issues require more future study?

We use these quesions to direct the review of the literature. We first examine the issues concerning different roles, and summarize the scenarios when collaboration and mutual understanding is required. Then we give some example patterns for cross-knowledge area collaboration. At last, we try to propose a routine by which the patterns of collaboration are evaluated and adapted for a given problem.

### B. Search Strategy

Our search strategy was set out to find the conjunction of requirements engineering, data science and machine learning. We conduct a search string-based database search on two specific digital libraries, IEEExplore and ACM Digital Library. For preventing from missing related papers, we use as few as words to filter the papers. We use *requirements* as a required word in title, while *requirements engineering* and *machine learning* are required as the author keywords of the search. The search is conducted by AND-operators. The year range from January 2016 to June 2022 is also adopted since we focus on the research that follows the recent trend of machine learning.

### C. Inclusion and Exclusion Criteria

The above search strategy yield 83 papers, 42 from IEEExplore and 41 from ACM Digital Library. We first executed our exclusion criteria over these papers. By our exclusion criteria, we filtered out the publications whose topic has less association with software engineering. An efficient way to do this is to filter out the papers whose title contains words like *teach*, *student*, *education* and *child*. A large number of the papers using machine learning to promote requirements engineering steps (commonly known as ML for RE) should also be filtered out because their motivations are not consistent with our research goals. We found that some words in the titles could help us locate them, like *automatic elicitation*, *automated identification*, *requirements classification* and *machine learning-driven requirements*. In addition to the above filtering methods, we had to complete the exclusion by reading the abstracts and checking the motivations. After executing the exclusion criteria, only 16 papers were left.

Then we conducted an iterative backward and forward Snowballing method for refining our results based on the remaining papers via Google Scholar. The scope was limited to software engineering methods for machine learning, machine learning applications, developement issues of machine learning ranging from 2016 to 2022. The final list of include 163 papers. The processes of filtering and refining were done by the first two authors, and a detailed discussion was held to reach consensus among all the authors.

## III. SURVEY RESULTS AND DISCUSSION

We first give out a list of all the selected papers in Table I. As an early milestone in the data-driven intelligence development paradigm, the Cross-Industry Standard Process for Data Mining (CRISP-DM) organizes related analytics activities into six phases: Business Understanding, Data Understanding, Data Preparation, Modeling, Evaluation and Deployment [168]. The CRISP-DM suggests a well-defined sequence of tasks with iterative feedback loops that suggests a requirements analysis cycle of data preparation, model design and evalution. Recently, CRISP-ML(Q) extends CRISP-DM to support the development of machine learning applications, whose special focus is on quality measurements of machine learning models, including robustness, scalability, explainability, model complexity and resource demands [169].

Vogelsang and Borg set out to define characteristics and challenges unique to Requirements Engineering (RE) for ML-based systems [20]. They identified several major changes in development paradigms, including the elicitation of ML performance measurements, the emerging of quality requirements such as explainability, freedom from discrimination, and specific legal requirements.

There are many recent proposals on software engineering approaches for machine learning applications. Amershi et al. [178] studied several representative example ML projects in Microsoft, in which several major challenges and success factors are summarised, including: sustainable end-to-end pipeline; data collection, cleaning and accessibility; model evaluation, evolution and deployment, etc. Then a nine-stage process model was proposed to address the above data-oriented challenges (e.g., collection, cleaning, and labeling) and model-oriented challenges (e.g., model requirements, feature engineering, training, evaluation, deployment, and monitoring), in which feedback loops are constructed from model evaluation and monitoring back to the previous stages, and from model training to feature engineering (e.g., in representation learning).

TABLE I: Topics of All the Seleted Papers

| Topics | Sum | Papers |
| --- | --- | --- |
| Big Picture | 15 | [10-24] |
| Stakeholders, Roles and Collaboration | 8 | [25-32] |
| Requirements Process Model | 7 | [33-39] |
| Requirements Elicitation and Specification | 9 | [3, 40-47] |
| Quality, Security, Ethics, and Assessment | 38 | [48-85] |
| Physics-Informed and Knowledge-based | 19 | [9, 86-103] |
| Machine Learning System Development | 15 | [103-117] |
| Interpretability and Explainability | 17 | [118-134] |
| Data Pipeline | 8 | [135-142] |
| Model Provenance, Verification | 7 | [143-149] |
| Applications | 18 | [150-167] |

TABLE II: Distribution of Requirements-Related Concerns for ML Applications

| Summary | Business Experts | Requirements Engineers | Software Engineers | Domain Experts | Data Scientists |
| --- | --- | --- | --- | --- | --- |
| Concerns (Functional Goals, Non-functional Requirements) | <ul><li>Business Goals</li><li>Accuracy</li><li>Stability</li><li>Efficiency</li><li>Fairness</li></ul> | <ul><li>Stakeholders</li><li>User Stories</li><li>Domain Models</li><li>Resources</li><li>System Scope</li></ul> | <ul><li>Prototyping</li><li>Architecture</li><li>Interface</li><li>Speed and Cost</li><li>Capacity</li></ul> | <ul><li>Mechanism design</li><li>Data Explanation</li><li>Knowledge acquisition</li></ul> | <ul><li>Data Pipeline</li><li>Task Definition</li><li>Train Resources</li><li>Model Performace</li><li>Explainability</li></ul> |
| Key challenges of RE for data-driven intelligence | In data-driven intelligent applications, the satisfaction of business goals are constrained by limitations of technological solutions. Sometimes the business experts have to make compromises and accept a less than expected solution. | The requirements process for data-driven intelligence applications is more complex than traditional requirements engineering, hence impose changes to existing vocabulary and requirements analysis tools. | The complexity of the software architecture requires extension to include data and machine learning models. What is more, it is harder to define the prototype which relies on a not unexplainable model. | Domain experts shares their understanding and knowledge about the working mechanism of a given problem. However, this is a progressive task as our understanding of the domain evolves constantly. | It is extremely challenging for data scientists as good quality data is always hard to get. Overcome this limitation and make good use of the available data, and convey technical limitations as early as possible are equally important. |
| Reference | [170] [171] [79] [47] | [80] [172] [173] | [17] [22] [174] | [175] [176] [87] | [20] [177] [32] |

Nalchigar et al. [39] proposes a modeling methodology representing generic ML design as solution patterns for business analytics. The pattern maps an actual business decision goal to a few questions, which are then answered through insights obtained from machine learning based on given data. Washizaki et al. [179] reviews architectural patterns and design patterns for ML systems covering different ML related tasks, such as datalake for storage, provision of raw data for analytics, decoupling of business logic from machine learning workflow, adoption of event-driven micro-services, version management of machine learning models, etc. The knowhow is rich and reusable but cannot cover ML application design process systematically. Trustworthiness of ML applications requires the compliance to applicable laws and regulations, as well as a series of domain specific physical laws. Hence the elicitation and evaluation of the compliance has become another major topic of interest in RE for ML. Sothilingam et al. [180] conducted an empirical case study of three ML software project organizations, and examined variations in project team designs using i* concepts of Agents, Roles, and Positions to support the analysis of complex organizational relationships for insufficient roles and expertises mapping.

There are related study on integrating scientific knowledge with machine learning for engineering and environmental systems, as well as hybrid modelling approaches that combine machine learning and simulations [181]. The integration could go both ways, either using ML to enhance domain models where the cause-effect relations are not fully evident [182], or using common-sense knowledge, common knowledge and domain knowledge models to modify generic models for specific domain. This is also called physics-aware learning or informed machine learning [98].

## A. RQ1: What are the roles involved in engineering data-driven intelligence applications?

In requirements engineering for traditional software development, the main roles are business experts, software requirements engineers and development engineers. A general requirements process starts with defining the scope of the business problem, which identifies the stakeholders by establishing the extent of the work. The software requirements engineer further identifies the requirements after requirements elicitation and specification through communication with the stakeholders, especially the business expert. When it comes to requirements of machine learning (or data-driven intelligence) functionalities, data scientists will take part in the RE process, and domain experts also play an irreplaceable role in industrial applications since domain knowledge are always necessary for understanding relevant theory and scenarios.

We summarize the concerns and challenges in process of RE for ML in Table II. It is not an exhaustive list, but include the ones that are most mentioned in the literature related to data-driven intelligence requirements. For example, fairness is introduced into the non-functional requirements since machine learning models can be biased by chosing training datasets in favor of certain group. And stability becomes more important than ever as the predictive results generated by machine learning models are unreliable when there is minor changes of situation.

The challenges stand for urgent problems to be solved from each role. For business experts, building a reasonable cognition on related technologies is quite meaningful, which would give the proposed business goals more supports. For requirements engineers, researchers have proposed some novel requirements modeling methods for machine learning applications in recent years, considering factors like privacy [183], security [76], scenarios [47] and goal revision [184]. For the other roles, the challenges mainly come from multidisciplinary and technical bottlenecks.

## B. RQ2: What are the areas for the engineers to collaborate during requirements stage?

In RE, there are many proven practices for the elicitation, modeling, specification, verification and management of requirements. These include goal-oriented modeling and analysis of functional requirements using KAOS and non-functional requirements using NFR, actor-based analysis to organizational structures with iStar, and scenario-based description of use-system interactions with use cases and use stories. These approaches well apply to the requirements processes for current industrial applications. For one example, the Volere Reqirements Process [185] is generally applicable to any early requirements stage when we try to understand the business context, form a system design idea, and verify it.

However, as we discussed in section III-A, the concerns of each role have changed and more roles must be involved. Digging into the concerns of each role, we can see the connection between them. For example, the business goals from business experts should be fulfilled by the prototypes from development engineers, while the prototypes must correctly use the machine learning models from data scientists. We decribe the connections in Fig. 1, where the roles are represented by circles, and red lines highlight the analysis process of using data-driven ML approach to address a problem.

Here we list the most widely discussed collaboration-related issues covered by the references.

- What should be considered if we want to use machine learning models as expected? This issue covers a wide range, including the widely concerned topic, XAI (or trustworthy AI). The collaboration on this issue generally happens between requirements engineers and data scientists. [77]
- How can software architectures be designed to enable robust integration of machine learning models? This issue exists because there is a huge gap between software development technologies and data science. The architectures design considerations have to include data quality, uncertainty, privacy and so on. Obviously this belongs to the partnership of development engineers and data scientists. [85]
- How can the process of requirements analysis be adaptive to machine learning systems? Due to big gap between traditional software and machine learning systems, existing requirements methods have to be improved accordingly.

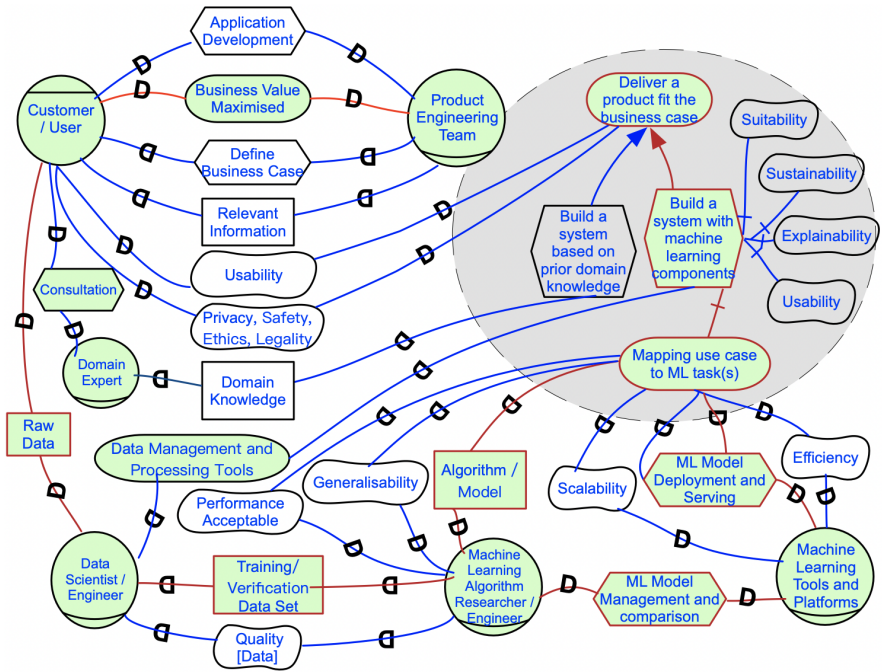

Fig. 1: General collaborations of Stakeholders involved in ML Application Development

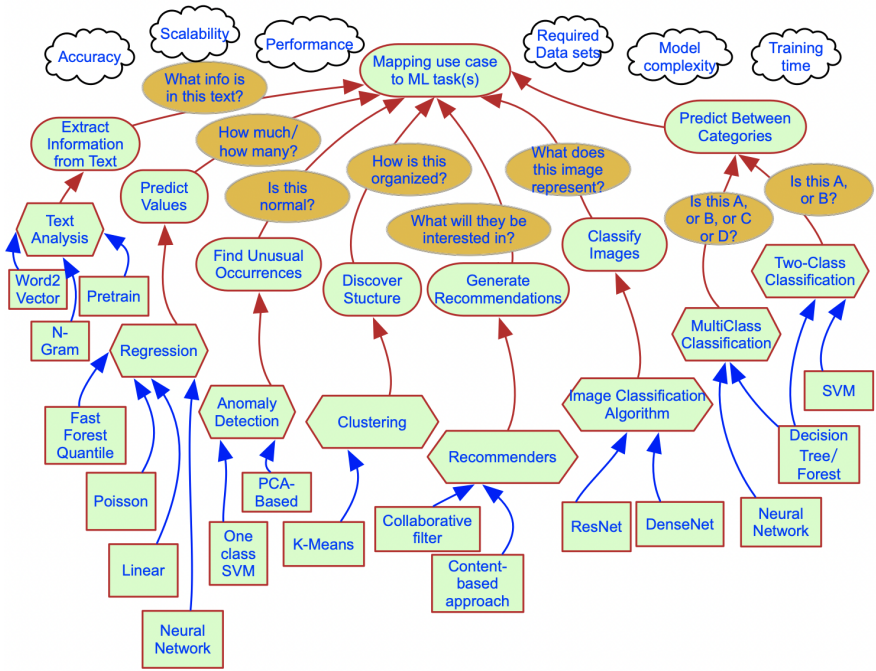

Fig. 2: Solution Mapping for Task Specific ML Applications

This issue is partly related to the above two issues, but from a higher perspective. [22]
- How can domain knowledge help design of machine learning models? The domain knowledge can be physical constrains, logic rules or knowledge graphs. To solve this issue, it requires close collaboration between domain experts and data scientists. [9]

*C. RQ3: What kind of support a collaborative requirements engineering for machine learning is needed?*

We present frameworks or patterns that are helpful for the collaboration in reuquirements engineering for machine learning from our selected papers. They are organized in the following two parts.

*1) Mapping Use Cases to Specific ML Tasks:* The process of data-driven intelligent system engineering requires several cross knowledge domain leaps: mapping a given use cases to corresponding machine learning task, building data pipeline and developing ML model, evaluating and deploying the model as software services. Nalchigar et al. [39] illustrate three solution patterns for machine learning that come from real world analytics projects in IBM. In each solution pattern, a concrete business goal is mapped to a business model with a specific machine learning task embeded, for which hierachical goal decomposition is conducted until an algorithmic solution is identified. Moreover, design rationale about how to develop a solution are represented as a context model showing the status of data, the motivations and technical constraints. The solution pattern provides an integrated view of multiple aspects of data-driven intelligent requirements or design decision making, for which we need a stepwised guideline to pilot the designers run through the process. The development of a specific ML application is never trivial, which could fail for many reasons, such as, poor data conditions, improper hyperparameter setting, or lack of algorithm selection. Therefore, evaluation criteria for acceptability should be carefully defined, including performance metrics, confidence and robustness, training cost, etc. Fig. 2 provides an overview of the general guideline for mapping where the red lines stand for collaborations and the blue lines belong to data scientists.

*2) General Guidelines for Integrating Domain Knowledge with ML:* Domain knowledge plays a key role in traditional requirements engineering in the development of most industrial applications. When come to knowledge-based cases, domain experts are extremely important for requirements engineers to understand business contexts and targets. For example, in the field of safety engineering, there are many well-established practice and tools for the evaluation of potential harmful events, such as: Hazard and Operability Analysis (HAZOP), Failure Modes and Effects Analysis (FMEA), Failure Modes, Effects and Criticality Analysis (FMECA), Layer of Protection Analysis (LOPA), Fault-Tree Analysis (FTA) and Event Tree Analysis (ETA), also called Bow-Tie Analysis, What-if Analysis, etc. These are practical models being widely used in process engineering fields, such as chemical engineering, pharmaceuticals, and nuclear energy engineering. It has been attracting the attention of software engineering researcher and practitioners since the 90's. With these approaches in place, practitioners build information systems to evaluate, manage potential risks of accidents. In recent years, tool vendors are looking into the possible intelligence extensions to existing functionalities. In order to build practical ML applications, it often requires knowledge fusion from multiple sources, those come from prior domain knowledge, and those come from data. For a pure ML process, data is fed into the machine learning pipeline, and produce the final prediction result, solves the problem by a ML model. Hence, we need to find alternative ways to incorporate knowledge into this pipeline. Rueden et al. [98] provide a survey that describes how different knowledge representations such as algebraic equations, logic rules, or simulation results can be used for machine learning. Four directions of integration are proposed, including training data generation, hypothesis set definition, learning algorithm modification and final hypothesis checking. Specifically, more than 30 strategies of integrating different knowledge and machine learning are described. Typically, scientific knowledge can be used in the design of loss term of deep learning models as a strong constraints. And regularization term based on the graph Laplacian matrix can enforce strongly connected variables to behave similarly in the model. We summarize this in Fig. 3.

*D. RQ4: What are the important issues require more future study?*

Requirements engineering for machine learning could be answering different questions for different people under different context. Depending on the roles or perspectives, the requirements to be elicited and analysed are different. As shown in Fig. 1 e.g. data scientists' main objective is to prepare a useful dataset for a given task; machine learning algorithms researchers' objective is to design a good foundation model that is adaptable to as many applications as possible, while ML engineers' goal is to improve the performance of a model by fine-tuning or selection of hyperparameter for a targeted problem; for designers of machine learning framework and platform, providing a model zoo and efficient model management services is of the top priority; for a system engineer, integrating ML component with traditional information systems techniques to address end-user needs is the ultimate goal. Each area has its own challenges to be addressed and requires relevant skillset and knowhow.

Besides these well-explored issues mentioned above, the following topics could be studied further:

- How a requirements model can adapt to dynamic changing scenarios and connect with a sustainable active machine learning pipeline? This is important for ML applications to handle situations with real-world complexity.
- How can we effectively produce a reliable overall cost estimation of a given project? Cost estimation is indispensable in traditional software development. However, machine learning technologies bring obstacles and challenges for this task. Unexpected cost may emerge anywhere during the process, including data collection, training, serving and model modification.
- How can we implement simulation-based prototyping for the development of machine learning applications? Early detection of prototype problems is essential to control risks. An effective way is to simulate the environment of applications, by which the technical methods can be verified in situ with limited additional cost.

IV. COLLABORATIVE REQUIREMENTS ANALYSIS PROCESS

In this section, we give a summary to the collaborative requirements analysis process and discus the main motivations of the selected papers by Fig. 4.

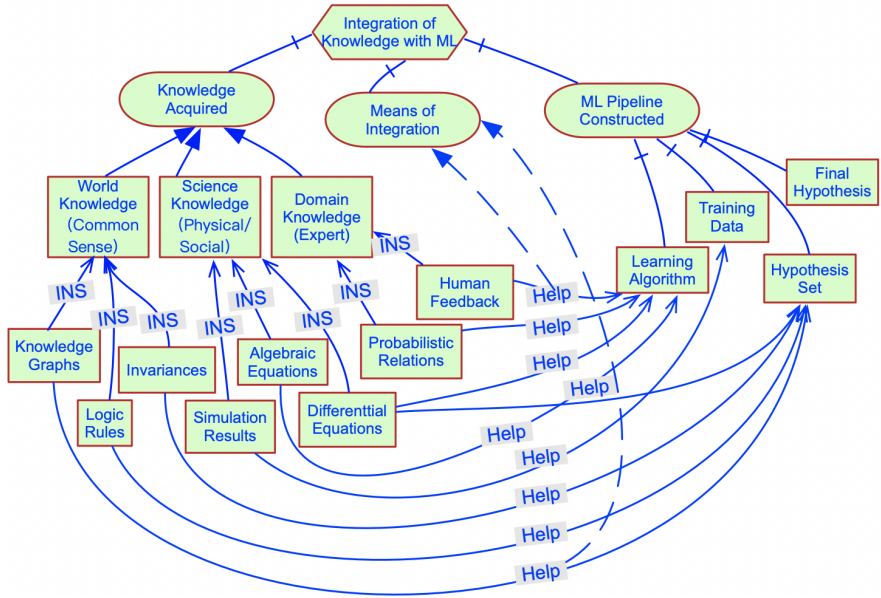

Fig. 3: A Reference Model for Integrating Knowledge with ML

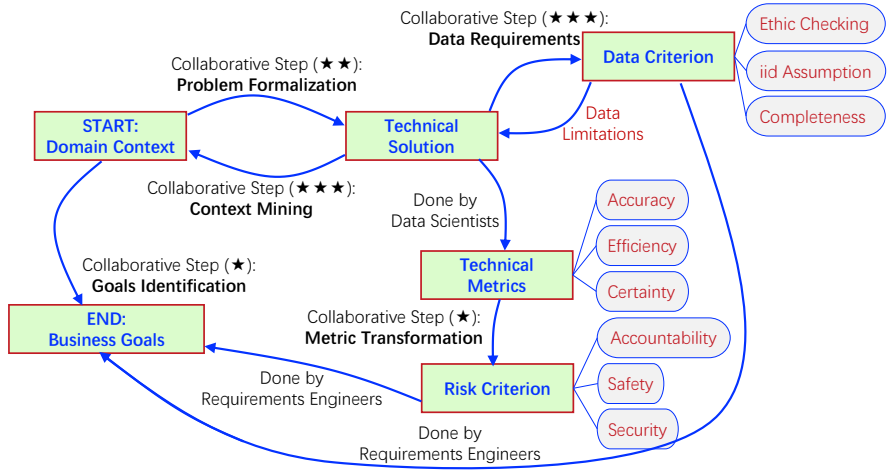

Fig. 4: Overview of Collaborative Requirements Analysis Process

### A. Collaborative Analysis Steps

In Fig. 4, we use green rectangles to stand for the inputs or outputs of the requirements analysis step. The blue arrowed lines means where the analysis happens, linking the input to the output. We use *Collaborative Step* with a few stars to mark the collaborative analysis steps, while the number of stars rates the complexity of a step. Except for the start node and the end node, the execution of the intermediate steps doesn't have to be in strict order. Here, we describe each step in detail:

- Problem Formalization: the step of problem formalization bridges the gap between business/domain experts and data scientists. However, in complex scenarios, formalizing requirements for data-driven intelligence is not that easy. Several types of formalization can be used, such as mathematical equations, logical rules and machine learning paradigm. After problem formalization, data scientists would match the problem with known solutions. This could be done by solution mapping as in Fig. 2. This step has a two star rating as domain knowledge is usually incomplete at the beginning, which makes it difficult to identify the technical problem right away.
- Context elicitation: A suitable technical solution requires efforts from business/domain experts and data scientists for understanding the situation. This step is not only data mining, but also mechanism mining and busniess logic mining. Details about this step is described in Fig. 3. This step has a three stars rating as it requires intensive interdisciplinary cooperation and there is no ready-made solution for it.
- Data Requirements Elicitation: To find a suitable tech-

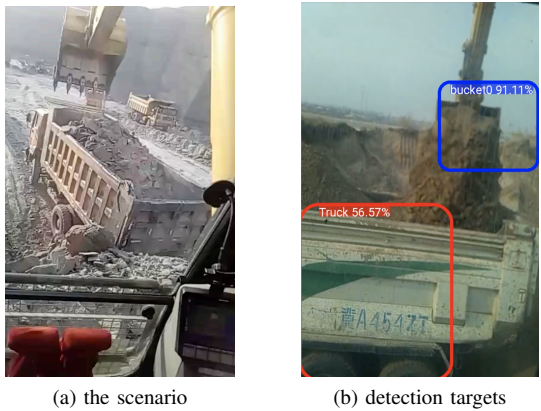

Fig. 5: A Virtual Excavator Supervisor with Smart Helmet

nical solution, data requirements have to be made clear. Business owners provide data source and examine for potential data ethics issues. Domain experts confirm that the data is used properly. Data scientists look closely to the completeness, sample distribution, iid assumption and so on. It is said that about 80% of time in a machine learning application development is comsumed for preparing data. Since coding is not that hard, what dominates the cost shoule probably be data requirements.
- Metric Translation: There is a gap between technical metrics and user understandable criterion. Data scientists and requirements engineers have to work together to translate the technical metrics to its business meaning.
- business goal evaluation: Finally, the business goals come from domain context, risk criterion and data criterion. Requirements engineers are responsible for making the business goals and values clear, and understandable to investors and end users.

### B. An Example Case

In this section, we discuss an example case, which is a retrospective study to a machine-learning application development case in relation to the questions discussed above. The business background will be introduced first, and then go through the collaborative requirements analysis steps. Through the example, we try to provide concrete evidences for why collaboration is needed and how hard it can be in real world.

**Business Case Description** The case is a virtual excavator supervisor application. In this use case, we were asked to develop a smart device based solution to replace the human supervisors of excavators' field work. The main task is to count the workload of machinary operators in terms of buckets of materials have been picked up and loaded into trucks, as shown in the images in Fig 5a. For an excavator leasing company, it is very important to track the workloads of each leased excavator. In the past few years, this task imposed high annual labor cost, which is expected to be replaced by automated solutions.

**Problem Formalization and Context Mining** Modern machinaries are often equipped with various preinstalled sensors. Our first option was to build a rule-based function to recognize the movements of buckets directly by analyzing signals collected from pre-installed sensors on the steel arms. However, as we could hardly tell the difference between discharging or excavating by reading the pressure signal, it is difficult to recognize a complete conveyance cycle and give an accurate count for workload. Inspired by the excellent performance of ML in computer vision tasks, we tried to analyze the bucket movement by analysing the images collected with a camera installed in front of the wind shield of the excavator. The operations of buckets are recorded as stacks of video files from which number of conveyances are expected to be recognised automately. However, there is no ready-made video image analysis algorithms directly usable for this task, as it is not a straightforward application of existing ML algorithms, such as object identification or posture recognition, etc. What made it even worse was the unreliable prediction results of the machine learning solution. We must pay good effort to collect data and ensure its quality. This task became a burden of the software engineers and data scientists, while sometimes domain exports have to provide professional and essential advices.

**Data Requirements Elicitation** The data is collected with a digital helmet. The first challenging data requirements is how to define the annotation rules. Different people have very different understandings about annotation of the digging buckets. For example, some one draws rectangles containing both the buckets and the stones in it, while others may prefer rectangles covering only the buckets. The differences will seriously affect the outcome of ML model training. Another common issue is unbalanced distribution of samples among data classes. The target with a small number of samples cannot be recognized with high confidence. The diversity of data and the clarity of images will bring challenging data requirements too. In practice, we can only cover a few cases of possible scenarios. Data availability, complexity of real world situation, and generalisability of machine learning models limit the practicality of machine learning-based solutions. Also, the overwhelming efforts required for data processing and data quality improvement is a last straw. Many ML application projects fail due to the poor generalisability of the model when facing new scenes, which may need better data requirements analysis.

**Metric Translation** When comparing the alternative machine learning algorithms of detecting buckets and trucks, a key criterion is Intersection over Union(IoU). However, this metric does not reflect the performance of counting the workload. For users, we need metrics like true positive rate and false positive rate.

**business goal evaluation** When setting business goals, we have to constrain the scenarios by the training set. True positive rate and false positive rate are required in existing scenarios. Besides recognition recordings should be remained for possible manual examination. The trained and verified model can still fail after deployment. Because there are often data shifts in real world applications, especially for the data from complex formative factors. Continuous monitoring and timely update are essential in order to maintain satisfying effect. Monitoring goals in the long term are important in goal identification.

## V. Conclusion and Future Work

In requirements engineering for machine learning applications, data description, performance metrics, data quality and candidate solutions, have to be iteratively and repeatedly orchestrated under a unified motivation. Failing in any single step can lead to the failure the entire project. In this paper, we provide an overview and reflection the collaboration among the different roles in requirements engineering for machine learning applications. We focused on the collaboration issues among business experts, requirements engineers, development engineers, domain experts and data scientists, including the integration of domain knowledge and machine learning models, how to use machine learning model as expected and so on. We further summarize the work that can be used to support collaborations, like the mapping from business cases to ML tasks, and practical reference of integration prior knowledge and machine learning workflow. An example cases of industrial data-driven intelligence applications are also provided.

Possible future work along the current line of research include: conducting more extensive empirical studies on success and failure cases industrial data-driven intelligence application projects; focus on the key issues identified and conduct more thorough case study; further evaluate the body of knowledge as (re)useable requirements and design patterns and form practical guidelines for effective collaborative requirements decision making on the alternatives ways for project success.

## Acknowledgment

Financial Support from the National key Research and Development Program Project 2021YFB1715200, and NSFC Innovation Group Project 62021002 is gratefully acknowledged. Collaborations with our industrial parnter TianYuan Technology is also acknowledged.